\documentclass[aps,prl,twocolumn,superscriptaddress,showpacs]{revtex4-1}

\usepackage {graphicx}
\usepackage{amsmath}
\usepackage{epstopdf}
\usepackage{color}

\def\b{\beta}

\begin{document}


\title{Compaction and tensile forces determine the accuracy of folding landscape parameters from single molecule pulling experiments}


\author{Greg Morrison}
\affiliation{School of Engineering and Applied Science, Harvard University, Cambridge, Massachusetts, 02138, USA}

\affiliation{Biophysics Program, Institute For Physical Science and Technology, University of Maryland, College Park, MD 20742}

\author{Changbong Hyeon}
\affiliation{School of Computational Sciences, Korea Institute for Advanced Study, Seoul 130-722, Republic of Korea}
\author{Michael Hinczewski}

\affiliation{Biophysics Program, Institute For Physical Science and Technology, University of Maryland, College Park, MD 20742}
\author{D. Thirumalai}
\affiliation{Biophysics Program, Institute For Physical Science and Technology, University of Maryland, College Park, MD 20742}


\begin{abstract}
We establish a framework for assessing whether the transition state location of a biopolymer, 
which can be inferred from  single molecule pulling experiments, corresponds to  the ensemble of structures that have equal probability of reaching either the folded or unfolded states ($P_{fold}$ = 0.5). 
Using results for the forced-unfolding of a RNA hairpin, an exactly soluble model and an analytic theory, we show that $P_{fold}$ is solely determined by $s$,
an experimentally measurable molecular tensegrity parameter, which is a ratio of the tensile force and a compaction force that stabilizes the folded state.  Applications to folding landscapes of DNA hairpins and leucine zipper with two barriers provide a structural interpretation of single molecule experimental data.
Our theory can be used to assess  whether molecular extension is a good reaction coordinate using measured free energy profiles. 
\end{abstract}

\pacs{87.10.-e,87.15.Cc,87.80.Nj,87.64.Dz}

\maketitle


The response of biopolymers to mechanical force ($f$), at the single molecule level, has produced direct estimates of many features of their folding landscapes, which in turn has given a deeper understanding of how proteins and RNA fold. 
In particular, single molecule pulling experiments directly measure distribution of forces needed to rupture biomolecules, roughness and shapes of folding landscapes \cite{Singlemol,Woodside06,Gebhardt10,Manosas06}.   Such measurements have made it possible to decipher the molecular origin of elasticity and mechanical stability of the building blocks of life, which is the first step in describing  how they interact to function in the cellular context. The major challenge is to provide a firm theoretical basis for interpreting the physical meaning and reliability of the folding landscape parameters that are extracted from trajectories that project dynamics in multi dimensional space onto one dimensional molecular extension, which is conjugate to $f$.

The key characteristics of the folding landscape of biomolecules that can be extracted from single molecule force spectroscopy (SMFS) measurements are  $f$-dependent position of the transition state (TS), the distance ($\Delta x^{\ddagger}=x_{TS}-x_{0}$) from the ensemble of conformations that define the basin of attraction corresponding to the native states (NBA), and the free energy barrier ($\Delta F^{\ddagger}$). 
The assumption in the analysis of SMFS data is that molecular extension is a good reaction coordinate for RNA and proteins, which implies that a single degree of freedom accurately describes the behavior of the multiple degrees of freedom explored by the biomolecule.  Structural meaning of $\Delta x^{\ddagger}$, a parameter that is unique to SMFS, has never been made clear. 
Despite many subtleties in determining $\Delta x^{\ddagger}$ from measurements \cite{Theory,HyeonTheory07,DudkoTheory06},  $x_{TS}$ is most easily identified as a local maximum of  the free energy profile at the transition mid-force 
$f_m$, $F(R|f_m)$, which can be constructed by measuring the statistics of end-to-end distance, $P(R)$ at $f=f_m$ \cite{Singlemol,Woodside06,Gebhardt10,Manosas06,HyeonRNAa,HyeonRNAb}.
This method has been experimentally used to obtain sequence dependent folding landscapes of DNA hairpins \cite{Woodside06}, and more recently proteins \cite{Gebhardt10}.   
In order to render physical meaning to $\Delta x^{\ddagger}$ 
we address two questions here: (1) Does $\Delta x^{\ddagger}$ describe the structures in the Transition State Ensemble (TSE)? The TSE describes a subset of structures that have equal probability of reaching the NBA or UBA staring from $\Delta x^{\ddagger}$. 
(2) Can a molecular tensegrity (short for tensional integrity) parameter \cite{IngberJCS03} $s$, expressing balance between the internal compaction force $f_m =\Delta F_{UF}/\Delta x_{UF}$ and the applied tensile force $f_c =\Delta F^{\ddagger}(f_m)/\Delta x^{\ddagger}(f_m)$($s=f_c/f_m$), describe the adequacy of $\Delta x^{\ddagger}(f_m)$ in describing the TSE structures?

We use simulations of a RNA hairpin and an exactly soluble model, both of which are apparent two-state folders as indicated  by $F(R|f)$,  to answer the two questions posed above.  
The TS is a surface in the multidimensional folding landscape (stochastic separatrix \cite{Pfolda}) across which the flux to the NBA and the Unfolded Basin of Attraction (UBA) is identical.  
This implies that the fraction of folding trajectories corresponding to  $\Delta x^{\ddagger}$ that start from the TS should have equal probability ($P_{fold}\approx 0.5$ \cite{Pfoldb}) of reaching the NBA and UBA \cite{Pfoldb,Pfoldc}. 
At $f=f_m$, the mean dwell times in NBA and the basin of attraction corresponding to unfolded conformations (UBA) are identical, so that $\int_{0}^{x_{TS}}dRP(R)=\int^{\infty}_{x_{TS}}dRP(R)=0.5$.  
However, it is unclear whether or not the barrier top position is consistent with the requirement $P_{fold} \approx 0.5$  in force experiments.

\begin {figure}
 \includegraphics[width=3.30in]{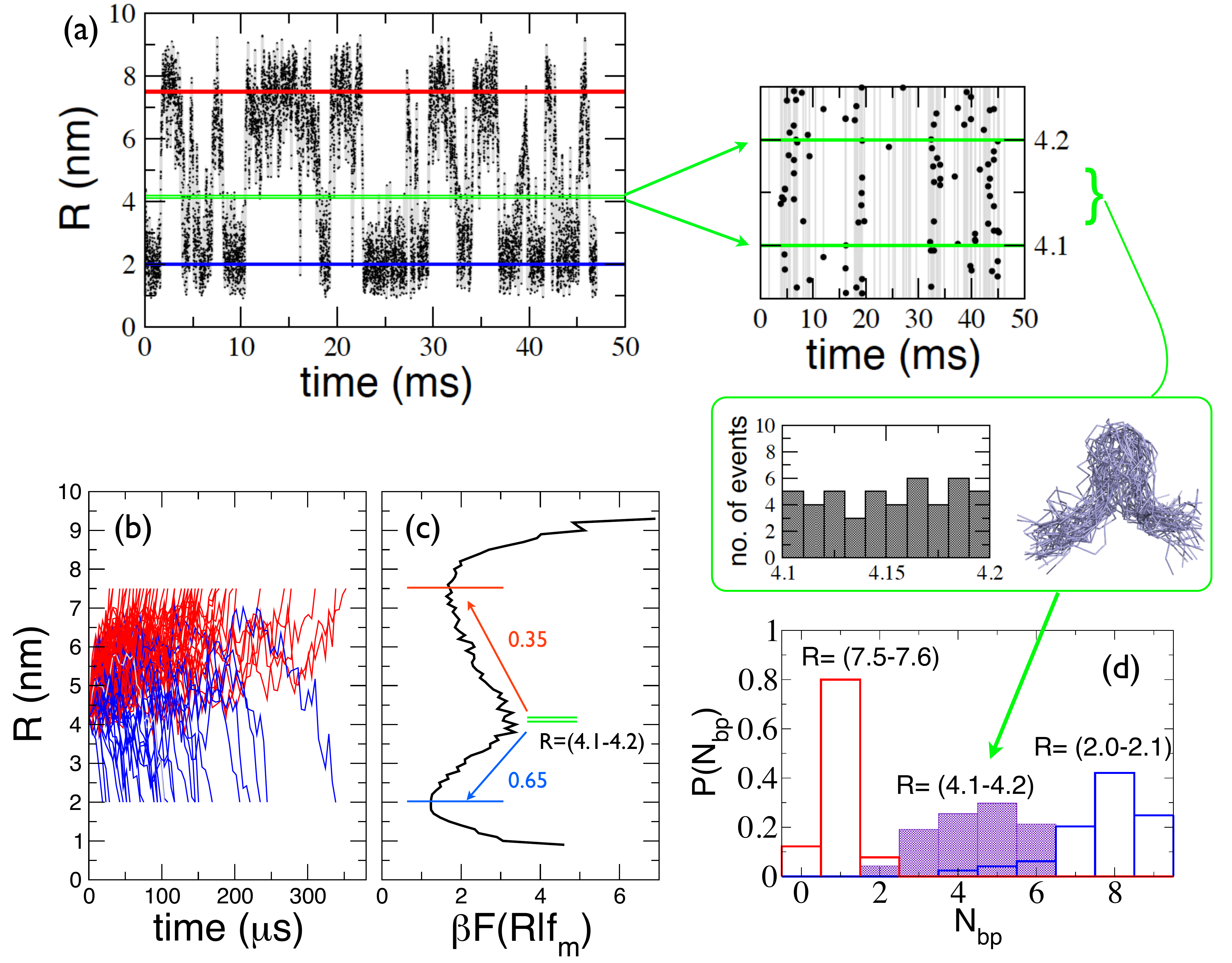}
 \caption{Relaxation dynamics of the ensemble of P5GA hairpins from the barrier top of $F(R|f_m)$. (a) $R(t)$ at $f=f_m$. TSE between $R=(4.1-4.2) $ nm is shown in the green box with a uniform R-distribution. 
 (b) Starting from the TSE of hairpins from (a), 
35 \% of the trajectories (red) reach UBA and 65 \% of the trajectories (blue) reach NBA. (c) Free energy profile at $f=f_m$. 
(d) TSE of P5GA has a broad distribution in the number of base pairs ($N_{bp}$).} 
\end{figure}


To ascertain,  whether the barrier top position is consistent with the requirement $P_{fold} \approx 0.5$  in force experiments,  we study  folding of P5GA, a RNA hairpin, for which the NBA and UBA are equally populated at $f_m=14.7$ pN \cite{HyeonRNAa,HyeonRNAb}.
 Both free energy profiles and the kinetics predicted by Kramers' theory show excellent agreement with the simulation results \cite{HyeonRNAb}, and formally establishes that extension $R$ is a good reaction coordinate for describing hopping kinetics at $f \approx f_m$.  
In practice, experimental time traces that have a  number of transitions as the one in Fig.1(a) can be used to estimate $P_{fold}$.  
With absorbing boundary conditions imposed at $R_N=2$ nm and $R_U=7.5$ nm (Fig.1(c)), we directly count the number of molecules from 47 points belonging to the TS region ($R=(4.1-4.2)$ nm (Fig.1(c)) that reaches $R<R_N$ (folded) and $R> R_U$ (unfolded). Although  the R-distribution of the 47 points is uniform (Fig.1(a))  we obtained $P_{fold}\approx 0.74$. 
To determine $P_{fold}$ using the ensemble method, we launched 100 trajectories from each of  the 47 structures and monitored their evolution (Fig.1(b)) using Brownian dynamics simulations \cite{VeitshansFoldDes97} with the multidimensional energy function for the hairpin \cite{HyeonRNAa}. We find that $P_{fold}=0.65$ (Fig.1(c)), which is similar to the value obtained by analyzing the folding trajectory.  
An examination of the individual trajectories  reveal that many molecules, initially with a gradient toward the UBA, re-cross the transition barrier to reach the NBA (blue trajectories in Fig.1(b)). 
Conversely, most of the molecules directly reach $R=R_N$ if they initially fall into the NBA, showing few recrossing events.
Although the precise percentage of molecules reaching UBA or NBA depends on the particular value of the boundary ($R_N$ and $R_U$), 
our simulations emphasize the importance of the \emph{re-crossing dynamics}, which is known to cause significant deviations from the transition state theory.  

Deviation from $P_{fold} \approx 0.50$ suggests that the global coordinate $R$ alone is not sufficient to rigorously describe the hopping kinetics of P5GA. At least one other auxiliary coordinate is needed, and for structural reasons we take it to be the number, $N_{bp}$, of base pairs \cite{Woodside06}. The broad asymmetric distribution of $N_{bp}$ within the narrow TS region ($R=(4.1-4.2)$) implied by $F(R|f_m)$ (Fig.1(d)) shows that hopping kinetics at $f_m$ should be described by multi (at least two) dimensional folding landscape even though the $f$-dependent rates of hopping between the NBA and UBA can be reliably predicted using $F(R|f_m)$ \cite{HyeonRNAb}. 

\begin {figure}
 \includegraphics[width=3.1in]{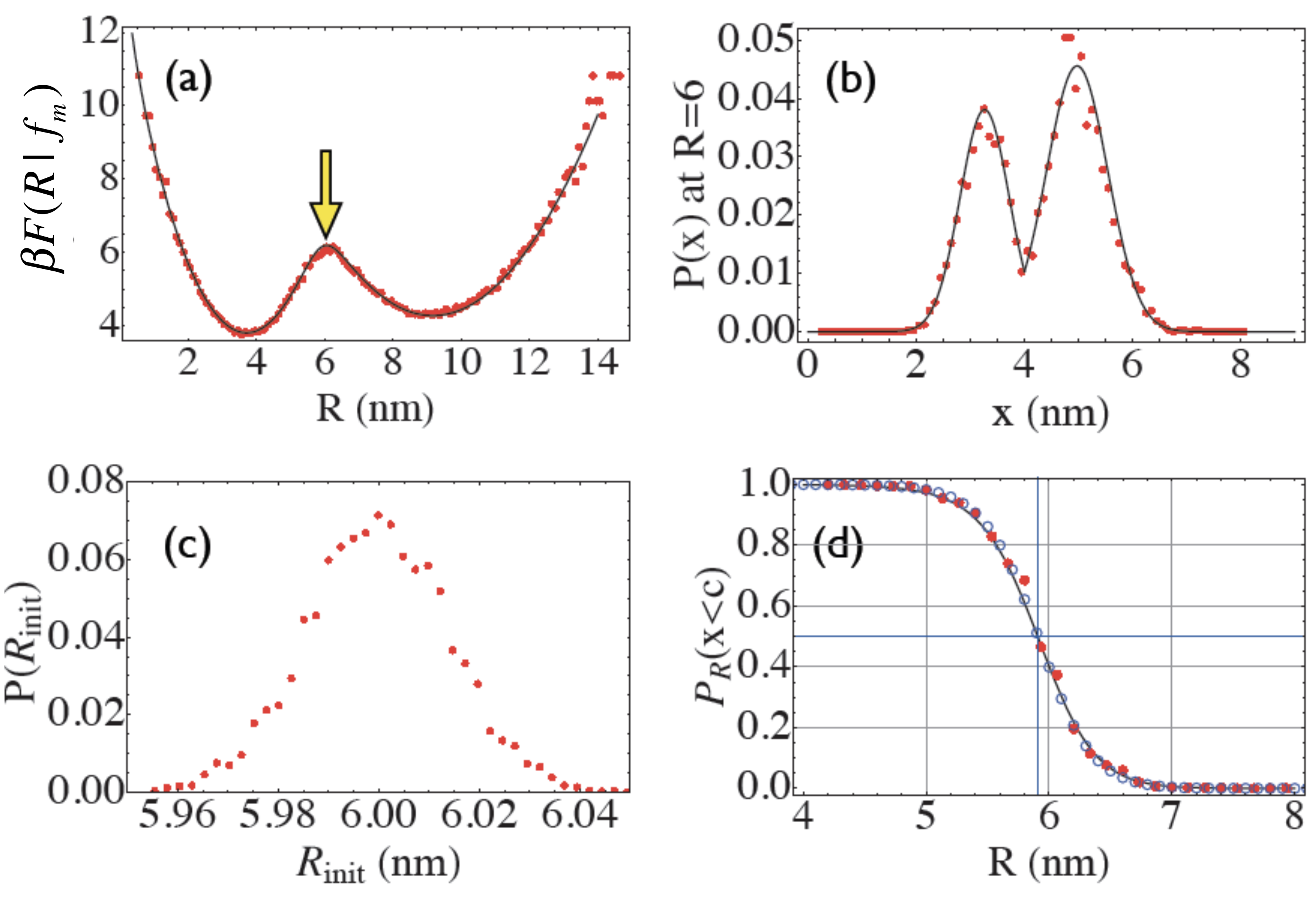}
 \caption{Relaxation dynamics of GRM chains from the barrier top of $\beta F(R|f_m)$. (a) $\beta F(R|f_m)$ for the GRM with interior interaction. Parameters are listed in the text. Red symbols are from the simulation result sampled using the chain with Hamiltonian of Eq.\ref{GRMdef}. Solid line is the theoretical fit. Arrow shows the position of barrier top. 
(b) The distribution of interior distance at $R=R_{TS}$. Black line is the theoretical prediction. 
(c) The initial distribution of $R$ prior to the relaxation dynamics, which is very tight round $R\approx R_{TS}=6$ nm. 
(d) The probability of being folded as a function of $R$. Open circles show the exact numerical solution for the sharp potential (Eq.\ref{GRMdef}). Filled circles are from simulations.  Solid black line shows the approximate solution using the smoothed potential (Eq.\ref{GRMsmooth}). 
}
\end{figure}

To further illustrate if $R$-coordinate alone is sufficient to determine the TSE structures we consider an analytically solvable Generalized Rouse Model (GRM) \cite{HyeonRNAb} that has a single bond in the interior of the chain whose presence corresponds to the NBA. 
The GRM Hamiltonian \cite{Barsegov08PRL}
\begin{align}
\beta \mathcal{H}&=\frac{3}{2a^2}\int^N_0ds\, \dot{{\bf r}}^2(s)-\beta {\bf f}\cdot\int^N_0ds\,\dot{{\bf r}}(s)\nonumber\\
&+\beta V_c\left[{\bf r}(s_1)-{\bf r}(s_2)\right],\label{GRMdef}
\end{align}
where $V_c[x]= k(x^2-c^2)/2$ for $x\le c$ and $V_c[x]=0$ for $x>c$,
describes a simple Gaussian chain under tension with an additional
cutoff harmonic interaction at the interior points $s_1$ and $s_2$.
The distribution of $R$ for the GRM, with $\Delta s = |s_2 - s_1|$, is
\begin{align}
&P(R;\Delta s)\propto R\sinh{(\beta f R)}e^{-3R^2/2(N-\Delta s)a^2}\times\nonumber\\
&\int^{\infty}_0dx x\sinh{\left[\frac{3Rx}{(N-\Delta s)a^2}\right]}e^{-3x^2N/2\Delta s(N-\Delta s)a^2-\beta V_c[x]}.
\end{align}
To obtain $P_{fold}$ we set $N=22$ for the number of bonds, $a=0.545$ nm for their spacing, $\beta=1/k_B T$, $\b k=1.65$ nm$^{-2}$, $c=4$ nm  for the strength and cutoff distance of the interior bond interaction, respectively, and $\Delta s=18$. For this set of parameters, the transition mid-force (where $P(x<c)=P(x>c))$ is $f_m\approx 16.8$ pN.   
The two minima and the position of barrier top of the free energy $\beta F(R|f_m)=-\log[P(R)]$ are easily determined as $R_N\approx 3.91$ nm, $R_U\approx 9.39$ nm, and $R_{TS}=6.00$ nm, respectively (Fig.2(a)). 

To calculate $P_{fold}$, we prepared 5000 GRM chains with $R=R_{TS}=6$ nm that corresponds to the maximum in $F(R|f_m)$ (Fig. 2a), and allowed the system to relax to either of the two basins of attraction, ending the simulation when the chain extension attains the value $R_N$ or $R_U$.  These simulations are performed using the Hamiltonian in Eq.\ref{GRMdef}, and not on the simple one-dimensional profile $\beta F(R)=-\log[P(R)]$.  
We find 60 \% (40 \%) of the initial chains from $R_{TS}$ reach $R=R_N$ ($R=R_U$), which implies $P_{fold} = 0.6$.  
Similar to the P5GA, the GRM dynamics projected onto the $R$-coordinate using both sets of parameters exhibit a number of recrossing events.  

To understand the relation between $x$ (the structural coordinate which specifies the NBA and UBA in the GRM)  and chain extension $R$ (pulling coordinate that is conjugate to $f$) we approximate the sharp interaction in Eq.\ref{GRMdef}  by a smoothed potential,  
\begin{align}
\beta V_c[x]\approx \beta V_S[x]=-\log\left(e^{-\beta k(x^2-c^2)/2}+1\right).\label{GRMsmooth}
\end{align} 
by taking advantage of the clear separation between the UBA and NBA (Fig. 2a). 
Defining $\mathcal{N}_k=\langle e^{-\beta k(x^2-c^2)/2}\delta (|{\bf R}|-R)\rangle_0$, with $\langle \cdots\rangle_0$ denoting an average over the Gaussian backbone, we can compute approximately $P_R(x<c)=\int_0^c dx\langle\delta(|{\bf R}|-R)\rangle\approx \mathcal{N}_k/(\mathcal{N}_{k=0}+\mathcal{N}_k)$, with $\langle\cdots\rangle$ representing an average over the GRM potential in Eq.\ref{GRMsmooth}.  The probability of bond formation satisfies
\begin{equation}
P_R(x<c)\approx\left[1+\lambda^{\frac{3}{2}}\exp{\left(\frac{3R^2\kappa\sigma^2}{2Na^2\lambda}-\frac{\beta kc^2}{2}\right)}\right]^{-1}\label{GRMprob}
\end{equation}
where $\kappa=Na^2\beta k/3$, $\sigma=\Delta s/N$, and $\lambda=1+\sigma(1-\sigma)\kappa$. 
At $R=6$ nm, Eq.\ref{GRMprob} gives $P_R(x<c)\approx 0.402\neq 0.5$. 
Thus, when the dynamics is initiated from the top of the apparent free energy barrier using the $R$ variable, their internal coordinate ($x$) is populated primarily with unfolded conformations (Fig.2(b)).  Despite the fact that the UBA is primarily populated at $R=R_{TS}$, we find that $P_{fold}\approx 0.6$, so that the NBA will be predominantly populated as the trajectories progress.  
The midpoint of transition  ($P_R(x<c) = 0.5$) occurs at  $R_{mid}=\left[(Na^2\lambda/3\kappa\sigma^2)(\beta kc^2-3\log{\lambda})\right]^{1/2}=5.91$ nm, which deviates slightly from $R_{TS}$ = 6.00 nm. 
The results from $V_S[x]$, which are in excellent agreement with the simulations as well as the numerical results using $V_c[x]$ (Fig.2), also suggests that hopping kinetics in the GRM  involves coupling between $x$ and $R$. Thus, even in this simple system accurate location of the TSE should consider two dimensional free energy profiles (see \cite{HyeonTheory07} and \cite{Suzuki10PRL}).

\begin {figure}

 \includegraphics[width=3.20in]{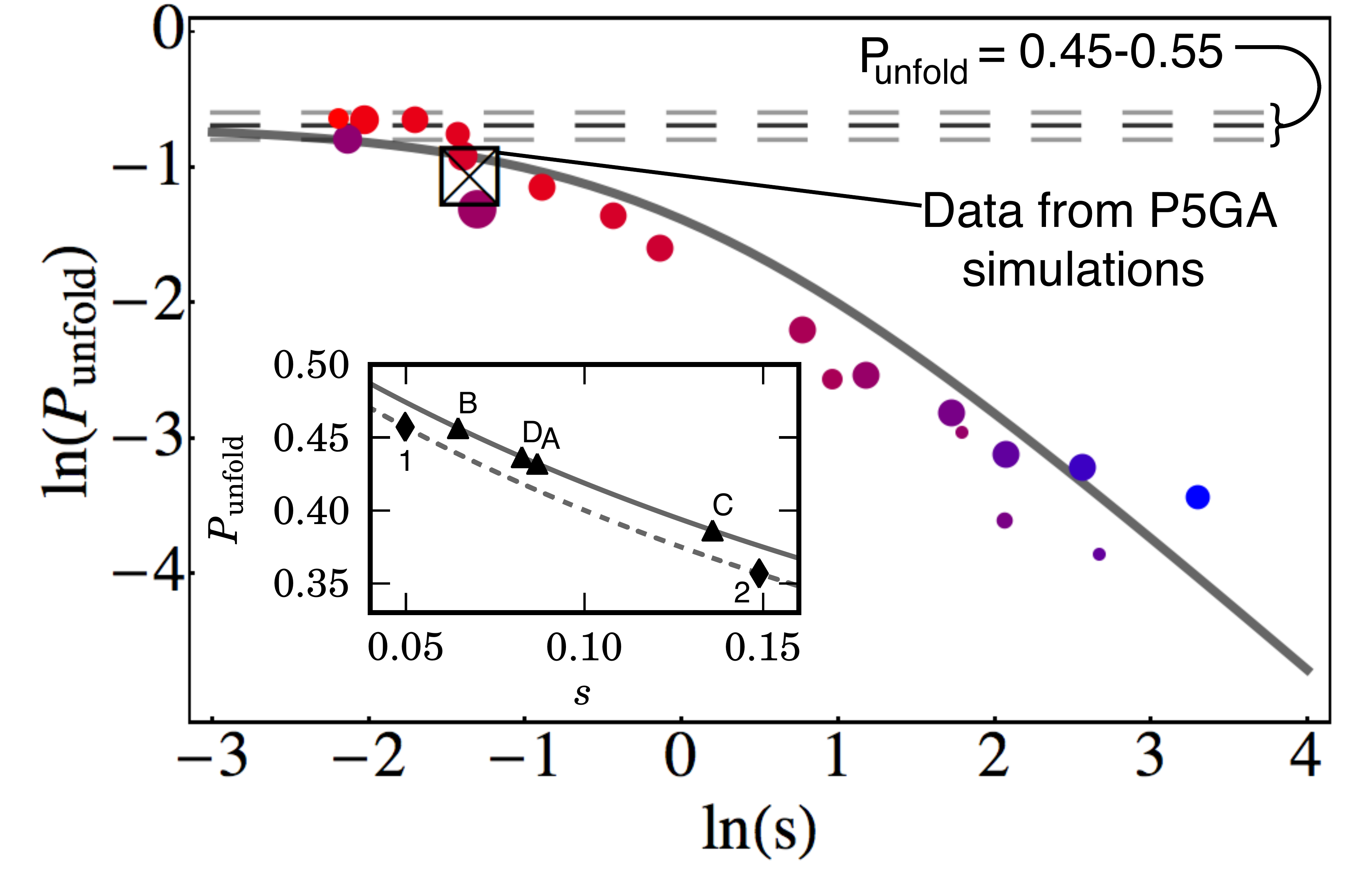}
 \caption{
 $P_{unfold}$ as a function of $s=\Delta F^ \ddagger/f_m\Delta x^ \ddagger $ for a variety of GRM parameters (colored circles).  Red corresponds to small $\beta k\sim (0.1-1)nm^{-2}$, purple to intermediate $\beta k\sim (1-50)nm^{-2}$, and blue to high $\beta k\sim (50-500)nm^{-2}$.  The point sizes indicate the magnitude of the midpoint force, with the smallest points at $f_m=5$pN and the largest at $f_m=25$pN.  The large square indicates the results of the P5GA simulations with $P_{unfold}=0.35$.  Gray line is the theoretical prediction $P_{unfold}(s)= (1+s)^{-1}/2$ for $s\gg 1$. Solid line in the inset shows the predicted $P_{unfold}(s)$ for $s\ll 1$ (Eq.5) using the DNA hairpin free energy profiles (not deconvolved) in \cite{Woodside06}.  Dashed line is for leucine zipper, which has two barriers, using data from \cite{Gebhardt10}. As explained in the Supplementary Information, numbers 1 and 2 are for NBA$\rightarrow$I and I$\rightarrow$UBA transitions, respectively. Because of different $k_U$ values (Eq.5) dashed and solid lines do not coincide.} 
\end{figure}

To answer the second question wes introduce a molecular tensegrity parameter, which is a ratio of tensile force and a force that determines the stability of  the biopolymer. The limit of mechanical stability of the NBA is determined by the critical unbinding force  $f_c=\Delta F^{\ddagger}(f_m)/\Delta x^{\ddagger}(f_m)$.  
At the midpoint force $f=f_m$, the tensegrity parameter  $s=f_c/f_m$ determines whether the applied external tension is sufficient to overcome the stability of the NBA. 
Models, which approximate the free energy profiles using a cubic or cusp potential \cite{DudkoTheory06}, could alter $f_c$ or $f_m$ from the form suggested above. However, because $s$ involves the ratio of the two forces the precise numerical factors are not relevant.
Barrier crossings between the NBA over the TS are governed by the competition between $f_c$ and the applied force $f=f_m$.  For $f_c\gg f_m$, barrier recrossing from the NBA to the TS will be extremely rare, while if $f_c\ll f_m$, barrier crossing events will be common.  We would therefore expect as the ratio $s=f_c/f_m$ increases, barrier recrossings from within the NBA to the TS will decrease, and the probability of reaching $R=R_{N}$ to increase.  Thus, $P_{fold}$, the structural link to $\Delta x^{\ddagger}$ should be determined by the experimentally measurable tensegrity parameter $s$. 

To confirm this expectation, we ran 20 different sets of GRM parameters (1000 runs each), with $f_m$ ranging between 5-25pN, $\beta k$ ranging from 0.1 to 600 nm$^{-2}$, $c$ between 0.26 and 6 nm, and $\Delta s$ between $N/2$ and $N$.  
Fig. 3 shows that  $P_{unfold}=1-P_{fold}$ decreases monotonically with $s=f_c/f_m$.  The dependence  of $P_{unfold}$ on $s$ (Fig.~3) can be derived 
by considering diffusion in a 1D version of the GRM potential in
Eq.(1), $\beta U(x) = -fx + \frac{1}{2}k_ux^2 + V_c(x)$ with $k_u$ being the curvature in the bottom of the UBA.  Assuming constant
diffusivity, $P_{unfold} = \int_{x_N}^c dx^\prime\,e^{\beta
  U(x^\prime)} / \int_{x_N}^{x_U} dx^\prime\,e^{\beta
  U(x^\prime)}$~\cite{vanKampen}, where the positions of the
transition barrier, native, and unfolded wells are given by $c$,
$x_N$, and $x_U$ respectively.  For $s \gg 1$, we obtain 
 $P_{unfold} = (1+s)^{-1}/2$, with the
entire dependence on the energy landscape encoded in $s$.  The analytical result, plotted as a solid curve in
Fig.~3, captures the overall trend of the GRM and P5GA simulations. 

For $s\ll 1$, $P_{unfold}$ can be approximated as:
\begin{equation}
P_{unfold} = \frac{1}{2}\left[\frac{\Phi}{1+\Phi} + \frac{32 f_m^2 \left(\Phi -1 \right)}{\pi k_\text{U} \left(\Phi +1 \right)^3}s\right]^{-1},
\end{equation}
where $\Phi=4f_m/\sqrt{2\pi k_B T k_\text{U}}$ and $k_\text{U}=\partial^2 U/\partial x^2$ at $x=x_\text{U}$.  This
relation can be used to predict the degree to which extension is a
good reaction coordinate for DNA hairpins.  We  calculated $s$ from the measured energy
landscapes in \cite{Woodside06}, and obtained
$P_{unfold}$ using Eq.(5).  We took  $f_m \approx 12.5$ pN,
and $\beta k_\text{U} \approx 0.2$ nm$^{-2}$.   The predicted $P_\text{fold}$ inset in Fig.~3) varies between
0.54--0.61 for the  DNA hairpin sequences A--D, close to the ideal value
of 1/2, indicating that extension is a reasonable coordinate except possibly for sequence C, which has a T:T base pair (bp)  mismatch 7  bp from the stem.   The limitation of extension as a reaction coordinate for this sequence is  consistent with the observation that folding of sequence C has an intermediate as indicated by the three minima in $F(R|f_m)$ \cite{Woodside06}.  

To illustrate that the theory is applicable when there are multiple barriers \cite{Manosas06}, we analyze the data for leucine zipper which unfolds by populating an intermediate. In this case there are two tensegrity parameters, $s_1$ (=0.05) and $s_2$ (=0.15) (see Supplementary Information). The predicted $P_{unfold}$ values show that extension is  a good reaction coordinate  for the NBA$\rightarrow$I transition but is less for the I$\rightarrow$UBA transition.(Fig.~3).   
 Our results can be confirmed by analyzing long time folding trajectories $R(t)$ as long as multiple hopping events occur.  
 Because $\Delta x^{\ddagger}$ is $f$-dependent, it follows that tensegrity  can be altered by changing $f$. Thus, extension  may be a good reaction coordinate over a certain range of $f$ but may not remain so under all loading conditions. 


\medskip
This work was supported in part by the grants from National Research Foundation of Korea (2010-0000602) (C.H.) and National Institutes of Health (GM089685) (D.T.).

\begin{center}
{\Large\bf Supplementary Information}\\
\end{center}

\noindent {\bf  Molecular Tensegrity parameters and $P_{fold}$ for Energy Landscapes with two barriers}\\[0.5em]

\vspace{-0.25em}
Here we provide details of how the theory in the main text can be extended to systems with multiple barriers using GCN4 as an example. The equilibrium time series for the GCN4 leucine zipper system was
measured in a constant extension dual optical trap setup \cite{Rief10PNAS}.  We can
express the bead-bead separation along the stretching direction as
$x_0 + x(t)$, where $x_0$ is the average separation, and $x(t)$ the
instantaneous deviation from the mean.  Constructing histograms from
the $x(t)$ trajectory yields the probability distribution ${\cal
  P}_\text{exp}(x)$.  In order to calculate the tensegrity parameters, we need measured  free energy profiles in the 
constant force (CF) ensemble, which is obtained from,
\begin{equation}
{\cal P}^0_\text{CF}(x) = C_0\exp(\beta k_\text{trap} x^2/4) {\cal P}_\text{exp}(x),
\end{equation}
where $k_\text{trap} = 0.25$ pN/nm is the optical trap stiffness \cite{Rief10PNAS},
$\beta = 1/k_B T$, and $C_0$ is a normalization constant.  
The
resulting distribution ${\cal P}^0_\text{CF}(x)$ is in the constant
force ensemble at a tension equal to $f_0$, the average force in the
experimental trajectory.  To obtain the distribution ${\cal
  P}_\text{CF}(x)$ at $f = f_0 + \delta
f$, we use
\begin{equation}
{\cal P}_\text{CF}(x) = C \exp(\beta x \delta f ) {\cal P}_\text{CF}^0(x),
\end{equation}
where $C$ is another normalization constant.  The associated free
energy $\beta F(x) = -\ln {\cal P}_\text{CF}(x)$.  

\begin{figure}
\includegraphics[width=2.00in]{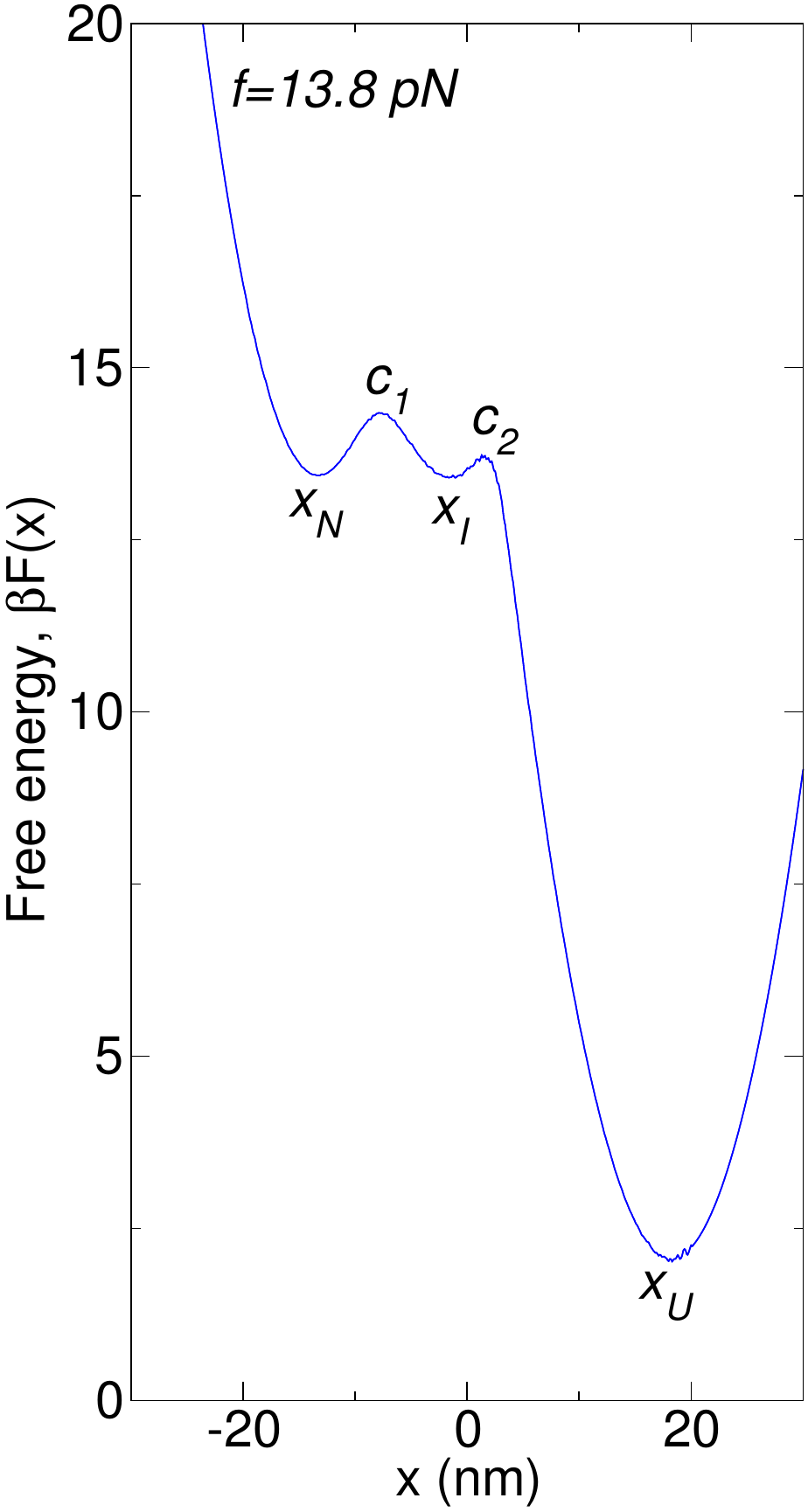}
\caption{Constant force free energy profile obtained using experimental data for leucine zipper (Fig. 3B in \cite{Rief10PNAS}) and Eqs. (6) and (7) at $f = 13.8$ pN, the midpoint of transition between NBA and I.}\label{fig1}
\end{figure}


 Fig.~\ref{fig1}
shows $\beta F(x)$  at  $13.8$ pN. The three states in the free energy landscape are NBA, an intermediate I, and UBA.
At $f=13.8$ pN, we see well-defined wells corresponding to
each of these states, with minima at $x_N$, $x_I$, and $x_U$
respectively.  At this force the probabilities of being in the NBA and I coincide *see below). The curvature $k_\alpha = (\partial^2 F/\partial
x^2)_{x=x_\alpha}$ of these wells, $\alpha = NBA$, $I$, $UBA$, can be
calculated numerically; the values are very similar for all three
wells, and so for simplicity we take
$k_\alpha \approx k \equiv 0.10$ $k_B T/\text{nm}^2$ (for both
$f=11.5$ and $13.8$ pN).  The transition barrier between NBA and I is at
position $c_1$, and between I and UBA at position $c_2$.

The first step in obtaining the tensegrity for each transition is to
determine the transition mid-force $f_m$.  We  estimate the probability
weight $P_\alpha$ associated with each well, assuming that the main
contribution comes from the region around the minimum.  In this
case,
\begin{equation}
P_\alpha \approx \sqrt{\frac{2\pi}{k}}\exp(\beta F(x_\alpha)).
\end{equation}
The mid-force condition for NBA to I yields a force $f_m = 13.8$ pN
where $P_{NBA} = P_I$.  Similarly the condition for I to UBA yields the
force $f_m = 11.5$ pN where $P_I = P_{UBA}$.  

The tensegrity parameters then follow from the shape of the free
energy landscape at each of these $f_m$:

\begin{center}\underline{Case I:  NBA to I, $f=f_m = 13.8$ pN}
\end{center}
\begin{equation}
\begin{split}
\Delta F^\ddagger &= F(c_1) - F(x_N) = 0.90\:k_B T\\
\Delta x^\ddagger &= c_1 - x_N = 5.4\:\text{nm}\\
s &= \Delta F^\ddagger/(f_m \Delta x^\ddagger) = 0.05
\end{split}
\end{equation}

\begin{center}\underline{Case II:  I to UBA, $f=f_m = 11.5$ pN}
\end{center}
\begin{equation}
\begin{split}
\Delta F^\ddagger &= F(c_2) - F(x_I) = 4.1\:k_B T\\
\Delta x^\ddagger &= c_2 - x_I = 9.8\:\text{nm}\\
s &= \Delta F^\ddagger/(f_m \Delta x^\ddagger) = 0.15
\end{split}
\end{equation}

Since both $s$ values are small, we use the $P_\text{unfold}$
expression for small $s$ to get the corresponding probabilities:
\begin{equation}
P_\text{unfold} = \frac{1}{2}\left[\frac{\Phi}{1+\Phi}+ \frac{4\Phi^2(\phi-1)}{(\Phi+1)^3} s \right]^{-1},
\end{equation}
where $\Phi = 4 \beta f_m /\sqrt{2\pi \beta k}$.  We plot the results
in the inset of Fig. 3 in the main text.  The dashed
curve corresponds to $f_m = 12.6$ pN (the average of 11.5 and 13.8
pN), since the difference between the two $f_m$ does not make a
noticeable difference in the predicted $P_\text{unfold}$ on the scale
of the figure.


\begin{thebibliography}{25}
\expandafter\ifx\csname natexlab\endcsname\relax\def\natexlab#1{#1}\fi
\expandafter\ifx\csname bibnamefont\endcsname\relax
  \def\bibnamefont#1{#1}\fi
\expandafter\ifx\csname bibfnamefont\endcsname\relax
  \def\bibfnamefont#1{#1}\fi
\expandafter\ifx\csname citenamefont\endcsname\relax
  \def\citenamefont#1{#1}\fi
\expandafter\ifx\csname url\endcsname\relax
  \def\url#1{\texttt{#1}}\fi
\expandafter\ifx\csname urlprefix\endcsname\relax\def\urlprefix{URL }\fi
\providecommand{\bibinfo}[2]{#2}
\providecommand{\eprint}[2][]{\url{#2}}

\bibitem[{\citenamefont{Singlemol}(2001)\citenamefont{Liphardt}}]{Singlemol}  
   \bibinfo{author}{\bibfnamefont{J.}~\bibnamefont{Liphardt et. al.}},
  \bibinfo{journal}{Science} \textbf{\bibinfo{volume}{292}},
  \bibinfo{pages}{733} (\bibinfo{year}{2001}); \bibinfo{author}{\bibfnamefont{B.}~\bibnamefont{Onoa et. al.}},
  \bibinfo{journal}{Science} \textbf{\bibinfo{volume}{299}},
  \bibinfo{pages}{1892} (\bibinfo{year}{2003}); 
   \bibinfo{author}{\bibfnamefont{C.}~\bibnamefont{Cecconi}},
  \bibinfo{author}{\bibfnamefont{E.~A.} \bibnamefont{Shank}},
  \bibinfo{author}{\bibfnamefont{C.}~\bibnamefont{Bustamante}},
  \bibnamefont{and} \bibinfo{author}{\bibfnamefont{S.}~\bibnamefont{Marqusee}},
  \bibinfo{journal}{Science} \textbf{\bibinfo{volume}{309}},
  \bibinfo{pages}{2057} (\bibinfo{year}{2005}); 
   \bibinfo{author}{\bibfnamefont{C.}~\bibnamefont{Hyeon}} \bibnamefont{and}
  \bibinfo{author}{\bibfnamefont{D.}~\bibnamefont{Thirumalai}},
  \bibinfo{journal}{Proc. Natl .Acad. Sci.} \textbf{\bibinfo{volume}{100}},
  \bibinfo{pages}{10249} (\bibinfo{year}{2003}); 
   \bibinfo{author}{\bibfnamefont{R.}~\bibnamefont{Nevo et. al.}},
  \bibinfo{journal}{EMBO reports} \textbf{\bibinfo{volume}{6}},
  \bibinfo{pages}{482} (\bibinfo{year}{2005}).

\bibitem[{\citenamefont{Woodside, Anthony,
  Behnke-Parks, Larizadeh, Herschlag, and Block}(2006)}]{Woodside06}
\bibinfo{author}{\bibfnamefont{M.~T.} \bibnamefont{Woodside et. al.}},
  \bibinfo{journal}{Science} \textbf{\bibinfo{volume}{314}},
  \bibinfo{pages}{1001} (\bibinfo{year}{2006}).

\bibitem[{\citenamefont{Gebhardt}(2010)}]{Gebhardt10}
\bibinfo{author}{\bibfnamefont{J.~C.~M.} \bibnamefont{Gebhardt}},
  \bibinfo{author}{\bibfnamefont{T.}~\bibnamefont{Bornschl{\"o}gl}},
  \bibnamefont{and} \bibinfo{author}{\bibfnamefont{M.}~\bibnamefont{Rief}},
  \bibinfo{journal}{Proc. Natl. Acad. Sci.} \textbf{\bibinfo{volume}{107}},
  \bibinfo{pages}{2013} (\bibinfo{year}{2010}).

\bibitem[{\citenamefont{Manosas}(2006)}]{Manosas06}
   \bibinfo{author}{\bibfnamefont{M.} \bibnamefont{Manosas}}, \bibinfo{author}{\bibfnamefont{D.}~\bibnamefont{Collins}},
  \bibnamefont{and} \bibinfo{author}{\bibfnamefont{F.}~\bibnamefont{Ritort}},
  \bibinfo{journal}{Phys. Rev. Lett.} \textbf{\bibinfo{volume}{96}},
  \bibinfo{pages}{218301} (\bibinfo{year}{2006}).

\bibitem[{\citenamefont{Hyeon and Thirumalai}(2005)}]{Theory}
 \bibinfo{author}{\bibfnamefont{C.}~\bibnamefont{Hyeon}} \bibnamefont{and}
  \bibinfo{author}{\bibfnamefont{D.}~\bibnamefont{Thirumalai}},
  \bibinfo{journal}{Proc. Natl. Acad. Sci.} \textbf{\bibinfo{volume}{102}},
  \bibinfo{pages}{6789} (\bibinfo{year}{2005}); 
 \bibinfo{author}{\bibfnamefont{C.}~\bibnamefont{Hyeon}} \bibnamefont{and}
  \bibinfo{author}{\bibfnamefont{D.}~\bibnamefont{Thirumalai}},
  \bibinfo{journal}{Biophys. J.} \textbf{\bibinfo{volume}{90}},
  \bibinfo{pages}{3410} (\bibinfo{year}{2006});
 \bibinfo{author}{\bibfnamefont{O.~K.} \bibnamefont{Dudko et. al.}},
  \bibinfo{journal}{Proc. Natl. Acad. Sci.} \textbf{\bibinfo{volume}{100}},
  \bibinfo{pages}{11378} (\bibinfo{year}{2003});
 \bibinfo{author}{\bibfnamefont{H.~J.} \bibnamefont{Lin}},
  \bibinfo{author}{\bibfnamefont{H.~Y.} \bibnamefont{Chen}},
  \bibinfo{author}{\bibfnamefont{Y.~J.} \bibnamefont{Sheng}}, \bibnamefont{and}
  \bibinfo{author}{\bibfnamefont{H.~K.} \bibnamefont{Tsao}},
  \bibinfo{journal}{Phys. Rev. Lett.} \textbf{\bibinfo{volume}{98}},
  \bibinfo{pages}{088304} (\bibinfo{year}{2007});
 \bibinfo{author}{\bibfnamefont{R.}~\bibnamefont{Merkel et. al.}},
  \bibinfo{journal}{Nature} \textbf{\bibinfo{volume}{397}}, \bibinfo{pages}{50}
  (\bibinfo{year}{1999});
 \bibinfo{author}{\bibfnamefont{I.}~\bibnamefont{Derenyi}},
  \bibinfo{author}{\bibfnamefont{D.}~\bibnamefont{Bartolo}}, \bibnamefont{and}
  \bibinfo{author}{\bibfnamefont{A.}~\bibnamefont{Ajdari}},
  \bibinfo{journal}{Biophys. J.} \textbf{\bibinfo{volume}{86}},
  \bibinfo{pages}{1263} (\bibinfo{year}{2004}).

\bibitem[{\citenamefont{Hyeon and Thirumalai}(2007\natexlab{a})}]{HyeonTheory07}
\bibinfo{author}{\bibfnamefont{C.}~\bibnamefont{Hyeon}} \bibnamefont{and}
  \bibinfo{author}{\bibfnamefont{D.}~\bibnamefont{Thirumalai}},
  \bibinfo{journal}{J. Phys.: Condens. Matter} \textbf{\bibinfo{volume}{19}},
  \bibinfo{pages}{113101} (\bibinfo{year}{2007}{\natexlab{a}}).

\bibitem[{\citenamefont{Dudko}(2006)}]{DudkoTheory06}
\bibinfo{author}{\bibfnamefont{O.~K.} \bibnamefont{Dudko}},
  \bibinfo{author}{\bibfnamefont{G.}~\bibnamefont{Hummer}}, \bibnamefont{and}
  \bibinfo{author}{\bibfnamefont{A.}~\bibnamefont{Szabo}},
  \bibinfo{journal}{Phys. Rev. Lett.} \textbf{\bibinfo{volume}{96}},
  \bibinfo{pages}{108101} (\bibinfo{year}{2006}).

\bibitem[{\citenamefont{Hyeon and Thirumalai}(2007{\natexlab{b}})}]{HyeonRNAa}
\bibinfo{author}{\bibfnamefont{C.}~\bibnamefont{Hyeon}} \bibnamefont{and}
  \bibinfo{author}{\bibfnamefont{D.}~\bibnamefont{Thirumalai}},
  \bibinfo{journal}{Biophys. J.} \textbf{\bibinfo{volume}{92}},
  \bibinfo{pages}{731} (\bibinfo{year}{2007}{\natexlab{b}}).

\bibitem[{\citenamefont{Hyeon and Thirumalai}(2008)}]{HyeonRNAb}
\bibinfo{author}{\bibfnamefont{C.}~\bibnamefont{Hyeon}},
  \bibinfo{author}{\bibfnamefont{G.}~\bibnamefont{Morrison}}, \bibnamefont{and}
  \bibinfo{author}{\bibfnamefont{D.}~\bibnamefont{Thirumalai}},
  \bibinfo{journal}{Proc. Natl. Acad. Sci.} \textbf{\bibinfo{volume}{105}},
  \bibinfo{pages}{9604} (\bibinfo{year}{2008}).

\bibitem[{\citenamefont{Ingber}(2003)}]{IngberJCS03}
\bibinfo{author}{\bibfnamefont{D.~E.} \bibnamefont{Ingber}},
  \bibinfo{journal}{J. Cell. Sci.} \textbf{\bibinfo{volume}{116}},
  \bibinfo{pages}{1157} (\bibinfo{year}{2003}).

\bibitem[{\citenamefont{Klosek et~al.}(1991)}]{Pfolda}
\bibinfo{author}{\bibfnamefont{M.~M.} \bibnamefont{Klosek}},
  \bibinfo{author}{\bibfnamefont{B.~J.} \bibnamefont{Matkowsky}},
  \bibnamefont{and} \bibinfo{author}{\bibfnamefont{Z.}~\bibnamefont{Schuss}},
  \bibinfo{journal}{Ber. Bunsenges. Phys. Chem.} \textbf{\bibinfo{volume}{95}},
  \bibinfo{pages}{331} (\bibinfo{year}{1991}).

\bibitem[{\citenamefont{Du et~al.}(1998)}]{Pfoldb}
\bibinfo{author}{\bibfnamefont{R.}~\bibnamefont{Du et. al.}},
  \bibinfo{journal}{J. Chem. Phys.} \textbf{\bibinfo{volume}{108}},
  \bibinfo{pages}{334} (\bibinfo{year}{1998}).

\bibitem[{\citenamefont{Klimov}(2001)}]{Pfoldc}
\bibinfo{author}{\bibfnamefont{D.~K.} \bibnamefont{Klimov}} \bibnamefont{and}
  \bibinfo{author}{\bibfnamefont{D.}~\bibnamefont{Thirumalai}},
  \bibinfo{journal}{Proteins - Struct. Funct. Gene.}
  \textbf{\bibinfo{volume}{43}}, \bibinfo{pages}{465} (\bibinfo{year}{2001}).

\bibitem[{\citenamefont{Veitshans et~al.}(1997)\citenamefont{Veitshans, Klimov,
  and Thirumalai}}]{VeitshansFoldDes97}
\bibinfo{author}{\bibfnamefont{T.}~\bibnamefont{Veitshans}},
  \bibinfo{author}{\bibfnamefont{D.}~\bibnamefont{Klimov}}, \bibnamefont{and}
  \bibinfo{author}{\bibfnamefont{D.}~\bibnamefont{Thirumalai}},
  \bibinfo{journal}{Folding Des.} \textbf{\bibinfo{volume}{2}},
  \bibinfo{pages}{1} (\bibinfo{year}{1997}).

\bibitem[{\citenamefont{Barsegov et~al.}(2008)\citenamefont{Barsegov, Morrison,
  and Thirumalai}}]{Barsegov08PRL}
\bibinfo{author}{\bibfnamefont{V.}~\bibnamefont{Barsegov}},
  \bibinfo{author}{\bibfnamefont{G.}~\bibnamefont{Morrison}}, \bibnamefont{and}
  \bibinfo{author}{\bibfnamefont{D.}~\bibnamefont{Thirumalai}},
  \bibinfo{journal}{Phys. Rev. Lett.} \textbf{\bibinfo{volume}{100}},
  \bibinfo{pages}{248102} (\bibinfo{year}{2008}).

\bibitem[{\citenamefont{Suzuki and Dudko}(2010)}]{Suzuki10PRL}
\bibinfo{author}{\bibfnamefont{Y.}~\bibnamefont{Suzuki}} \bibnamefont{and}
  \bibinfo{author}{\bibfnamefont{O.~K.} \bibnamefont{Dudko}},
  \bibinfo{journal}{Phys. Rev. Lett.} \textbf{\bibinfo{volume}{104}},
  \bibinfo{pages}{048101} (\bibinfo{year}{2010}).

\bibitem{vanKampen} N.G. van Kampen, {\it Stochastic Processes in
  Chemistry and Physics}, 3rd ed. (Elsevier, Amsterdam, 2007),
  pg. 305.

\end{thebibliography}

\begin{thebibliography}{100}
\bibitem{Rief10PNAS}
Gebhardt, J.C.M.; Bornschl{\"o}gl, T. \& Rief, M.
\newblock {Full distance-resolved folding energy landscape of one single protein molecule}.
\newblock {\em Proc. Natl. Acad. Sci. USA} {\bf 107}, 2013.
\newblock (2010).
\end{thebibliography}

\end{document}